\documentclass[aps,amsmath,prb,a4paper,floatfix,twocolumn]{revtex4}
\long\def\comment#1{}
\usepackage{graphicx}
\usepackage{amsmath}
\usepackage{amsfonts}
\usepackage{amssymb}

\newcommand{\beq}{\begin{equation}}
\newcommand{\eneq}{\end{equation}}
\newcommand{\bea}{\begin{eqnarray}}
\newcommand{\enea}{\end{eqnarray}}

\newcommand{\fpr}{F$\pi$R~}
\newcommand{\pari}{$\mathcal{P}~$}

\begin{document}
\title{ Topological $rf-$SQUID   with a  frustrating $\pi-$ junction for  probing  the  Majorana Bound State}

\author{P. Lucignano$^{1,2}$}
\author{F.Tafuri$^{3,1}$}
\author{A. Tagliacozzo$^{2,1}$}

\affiliation{$^1$ CNR-SPIN, Monte S.Angelo -- via Cinthia,  I-80126 Napoli, Italy}
\affiliation{$^2$ Dipartimento di Scienze Fisiche, Universit\`a di Napoli ``Federico II'', Monte S.Angelo, I-80126 Napoli, Italy}
\affiliation{$^3$ Dipartimento Ingegneria dell'Informazione, Seconda Universit\`a  di Napoli, I-81031Aversa (CE), Italy}

\date{\today}

\begin{abstract}
Majorana Bound States are predicted to appear as boundary states of the Kitaev model. 
Here we  show that a $\pi -$Josephson Junction, inserted in a topologically non trivial  model ring,  sustains a Majorana Bound State, which is robust with respect to local and non local perturbations. The realistic structure could be based on a High Tc Superconductor tricrystal structure, similar to the one used to spot the  d-wave order parameter. The presence of the Majorana Bound State changes the ground state of the topologically non trivial  ring   in a measurable way, with respect to that of a conventional one. 
\end{abstract}

\maketitle
\section{Introduction}

After the pioneering theoretical proposal by Kitaev\cite{Kitaev:2001}, Majorana Fermions (MFs) have been predicted in a wide class of low-dimensional solid state devices. Being neutral excitations in Fractional Quantum Hall systems or  hybrid superconducting devices, MFs are highly attractive for quantum computing gates, as well as for fundmental reasons.   Despite the considerable theoretical and experimental efforts \cite{Alicea:2012}, challenges still remain before a real solid state device  can be realized, allowing for isolation and manipulation of MFs. Among the promising 
systems, there  are  superconductors in contact with topological insulators (TIs) \cite{Teo:2010} or quasi one-dimensional systems with strong spin orbit interactions  \cite{Lutchyn:2010,Oreg:2010,Duckheim:2011,Weng:2011,Pikulin:2012},  helical magnets \cite{Kjaergaard:2011} and other materials \cite{Alicea:2010,Fu:2009,Akhmerov:2009,Tanaka:2009,Linder:2010}.  However  several  issues have to be convincingly solved \cite{Potter:2011,Lutchyn:2011,Brouwer:2011} and the recently announced transport measurements \cite{Kouwenhoven:2012,Deng:2012,Heiblum:2012} still arise excitement and debate \cite{DeFranceschi:2012}. Disappointingly, their  zero bias anomalies can be  fitted by both MF physics and Kondo or 0.7 anomaly physics \cite{DeFranceschi:2012,Marcus:2013}. 

Adopting a different point of view, we leave aside transport measurements, and  in this paper we  propose a magnetic flux measurement on a device combining the physics of topological insulators and superconductive d-wave systems.
We show that the spontaneous flux generated in the ground state (GS) of a frustrated topological SQUID Josephson ring (we call it "frustrated $\pi-$ring" (\fpr) in the following) can be unambiguously  related to the presence of  MFs. 

Feasible realizations of such a system could be a semiconducting nano-wire with strong spin-orbit coupling (e.g. InAs or InSb), with  spin polarization splitted by a Zeeman  field,  or, alternatively  nano-wires made by 3D TIs (Bi2Se3 or Bi2Te3), in contact with a superconductor. Superconductors in contact with the edge states of a  2D topological insulator (HgTe) could be also considered \cite{FuKane:2009}.
The frustration is obtained employing, as superconducting material, a high Tc tricrystal structure, realized by epitaxial growing of high Tc (HTS) cuprates,  matching three differently oriented crystals.  Rings built on  tricrystals  have provided the evidence for the $d-wave$ pairing in HTS materials \cite{Tsuei:1994}.

 In Ref.~\cite{Tsuei:1994} the experiment has been properly designed so that the ring contains an odd number of $\pi$-junctions and has a total critical current ($I_c$) and an inductance ($L$) to guarantee $I_c L>>\phi_o$  (here $\phi_0=hc/2e$ is the superconductive flux quantum).  Its GS  spontaneously breaks the time reversal  symmetry with a current flowing in the ring and generating  a spontaneous  fractional flux ${\phi=\pm \phi_0/2}$ that can be measured by scanning SQUID microscopy \cite{Tsuei:2000}. 

\begin{figure}[!htbp]
\begin{center}
\includegraphics[width=0.49\linewidth]{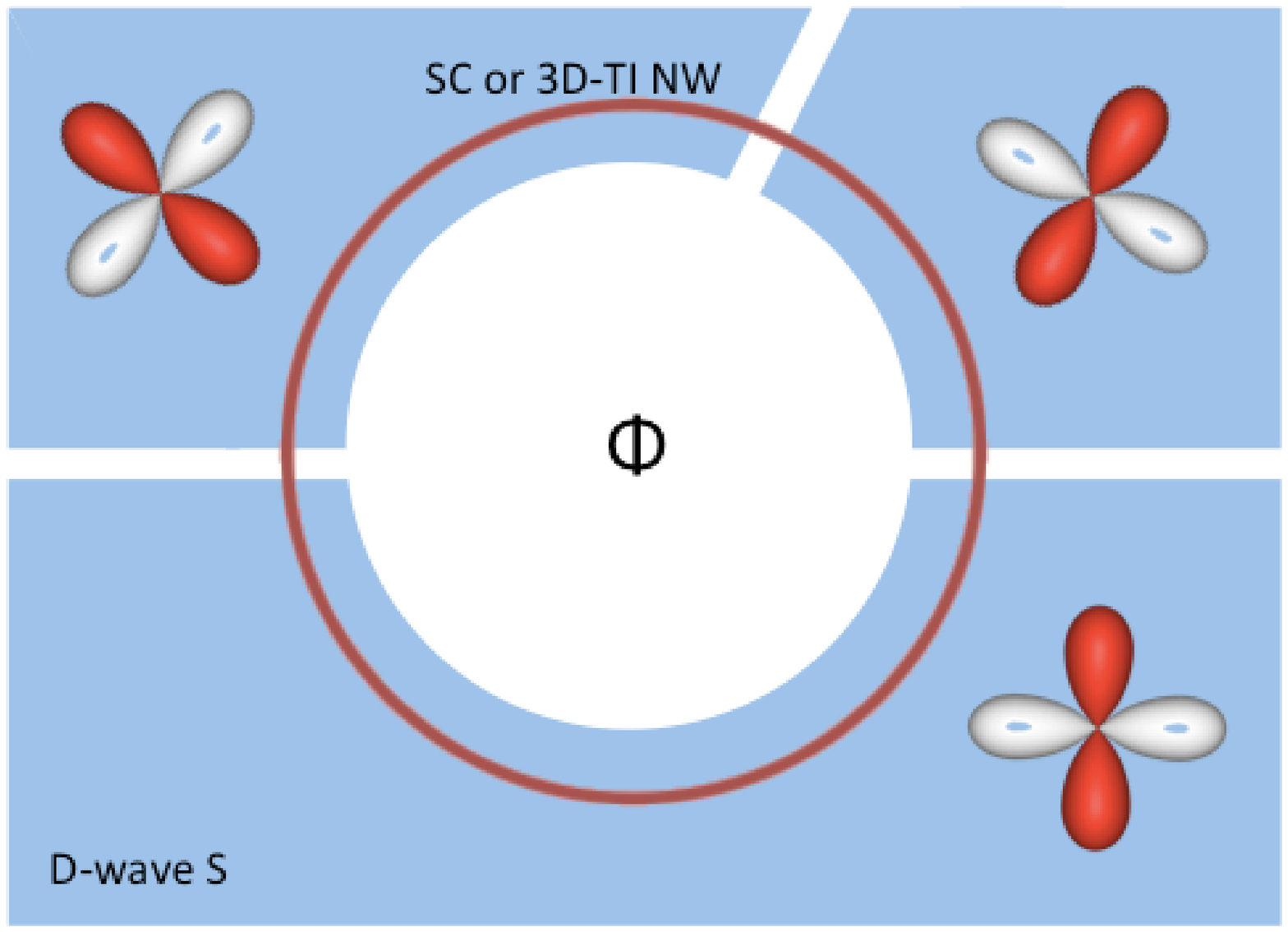}
\includegraphics[width=0.49\linewidth]{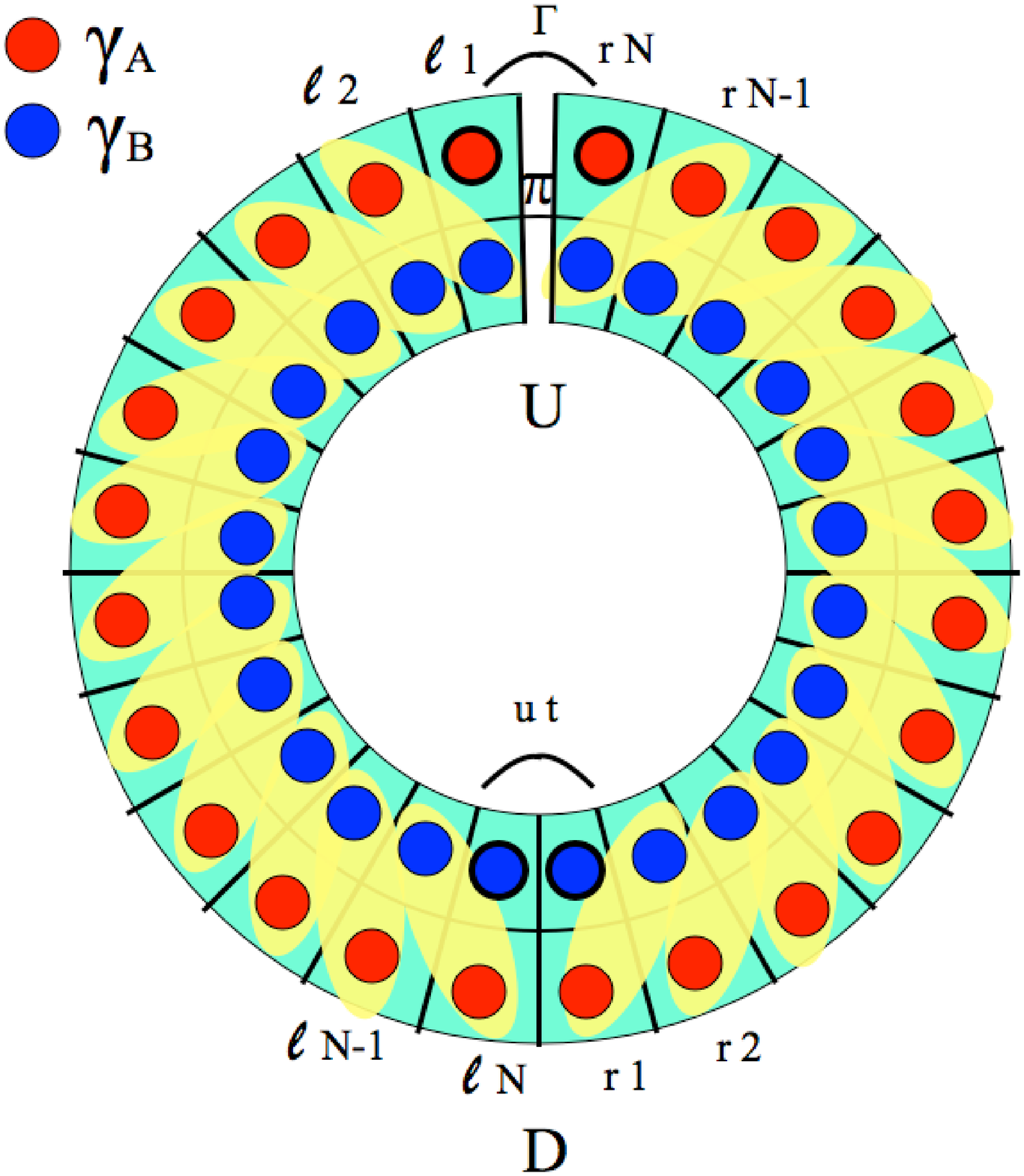}
\caption{(color online) Left Panel) A Ring made of an InAs or 3D TI nano wires is deposited onto a  high Tc tricrystal superconductor. Right panel) Sketch of the Kitaev chain with a $ \pi -$junction at the top point $U$. Red (blue) circles represent $\gamma_{A(B)}$ type MFs. Yellow ellipses signal strong coupling.  }
\label{fig1}
\end{center}
\end{figure}

Our proposed devices is sketched in Fig.{\ref{fig1}}.
Nano wires, either made out of semiconductors of 3DTI, mimicks quasi 1D system, properly described by the Kitaev model \cite{Alicea:2012}.  

%
%
Inspired by the concept that any superconducting loop with an odd number of $\pi$  junctions  has a frustrated GS \cite{Sigrist:1995}, we study a closed loop "Kitaev chain"  \cite{Kitaev:2001}, with one single weak link having a jump of $\varphi= \pi$ in the phase of the OP (see Fig.\ref{fig1} for a sketch).  

We show the various branches in the energy dispersion of the Andreev Bound states $E_n(\varphi )$ of the model ring\cite{Kitaev:2001}, finding a  crossing of states of opposite fermionic parities  \pari ~  at zero energy, which  signals unambiguously   the presence of the Majorana Bound State (MBS) in the spectrum.    

As shown in Ref.~\cite{Lucignano:2012}, $\pi$ junctions 
can display a MBS  at a phase difference  $\varphi=0$, rather than  $\varphi=\pi$. 
Here a corresponding state is found in the ring geometry, stabilized with respect to residual interactions.

The free energy describing a Josephson rf-SQUID ring shows  minima at a phase difference $\varphi$ corresponding to the  measurable spontaneous flux. In the conventional  \fpr geometry, there is quite a high barrier separating the two minima at  $\pm \phi _o/2$, which freezes the frustrated system in a  time reversal symmetry broken GS.  
In our model, we find two minima close to  $\pm \phi _o/2$ as well, as  in the conventional \fpr. However, they   correspond to different parities.  This marks a fundamental difference with respect to unfrustrated structures (conventional "0" rings), where the GS is non degenerate (at zero flux) and fixes the parity.

In our case, $\mathcal{P}$ conservation, would imply that changing the number of electrons (for instance by means of a quantum point contact) should affect the measured spontaneous flux, thus marking a strong difference w.r. to a conventional \fpr.
However, while the isolated model Hamiltonian conserves \pari, this symmetry is rather unlikely to be mantained in a real device, where impurities, can trap  and release charges. These events  would  correlate to flux quanta entering or leaving  the ring, thus inducing  tunneling of the system  between the two minima, so that the expected  average flux associated to the GS is $\langle \Phi \rangle = 0$. Measuring zero flux at a \fpr,  would be the smoking gun evidence for  the existence of  the MBS in this device.  It is a weird case  that the breaking of the discrete symmetry (\pari ) enforces the  time reversal symmetry to be restored. 

The paper is structured as follows. In sec. II we introduce the model Hamiltonian. In sec. III we show the energy spectrum of a Pi ring compared with the one of a Pi Junction. In sec IV we study the free energy landscapes depending on the fermion parity and discuss the possible spontaneous fluxes threading the ring. Conclusions are summarized in sec. V.
 
\section{Model Hamiltonian} 

We  consider a N-sites Kitaev chain  of unitary lattice spacing and  full length $L = 2 N $, with  real  inter site hopping $t$.  When folded in the shape of a ring the system displays mirror  symmetry  across the vertical line connecting  points U and D between  the  top and the bottom (see Fig.~\ref{fig1}). 
  
Our device can be described as two N-sites Majorana wires (left ($\ell$)  and right ($r $)),  coupled at the top of the ring $U$ by the weak electron tunneling of energy  $\Gamma$.  At chemical potential $\mu =0 $, in the presence of an electromagnetic vector potential $\vec{A}$, the gauge invariant Hamiltonian  reads as $H=H_{\ell}+H_{r}$, with: 
\begin{widetext}
\beq
H_{\alpha}= \sum _{ j=1}^{N-1} \left (   -\frac t 2   e^{ig_{\alpha j}} c^\dagger _{\alpha j} c_{\alpha j+1} +\frac  \Delta 2 \: e^{i(\phi _{\alpha j}+\phi_{\alpha j+1})/2}  \: c_{\alpha j}   \: c_{\alpha j+1} + h.c. \right ).
\eneq 
\end{widetext}
Here $\alpha $ labels the $\ell $ and $r$ side of the ring,  $c_ {\alpha j}$ are spinless Dirac fermions at the site $j$ and  $\Delta$ is the  effective  p-wave superconductive pairing.
The phases $g_{\alpha j}$ acquired in the hopping between the $j$ site and its nearest neighbor  and the gauge invariant phase $\phi_{\alpha j}$ are defined as:
\bea
g_{\alpha j} = -\frac{e}{\hbar c} \int_{\alpha j}^{\alpha j+1} \vec A \cdot d \vec l\:,\\
\phi_{\alpha j} = \theta_{\alpha j} - \frac{2e}{\hbar c} \int _ {\ell 1}^{\alpha j} \vec A \cdot d \vec l\:,
\enea
where $\theta_{ \alpha j} $ is the phase of the superconducting order parameter.



  At the top point  U of the ring (see Fig.1) there is a tunnel junction:   %
\bea
H_{U} = -\Gamma\:  \left(c^\dagger_{\ell 1} c_{rN} + h.c. \right)\:,
\enea
($\Gamma << t $). 
For sake of further investigations, we explicitly consider also the hopping term at  the bottom point in the ring  D, where the $\ell $  and $ r $  chains are matched:
\bea
H_{D} = -u\:  t \:  \left(c^\dagger_{\ell N} c_{r1} + h.c. \right)\:.
\enea
Here we will keep the dimensionless parameter  $ u $ (which may be complex) as a variable,  to discuss also the trivial case of   $u=0$, which  corresponds to the  ring  cut at D,  with open ends. 

The spinless Dirac fermions can be expressed in terms of  two species of Majorana fermions at each site of the ring $\gamma_{A/B j}^{\alpha}$,  such that:
\bea
\gamma^\alpha_{Bj} = c_{\alpha j} e^{i \phi_{\alpha j}/2}+ c_{\alpha j}^\dagger e^{-i \phi_{\alpha j}/2}\:,\\
\gamma^\alpha_{Aj} =-i \left ( c_{\alpha j} e^{i \phi_{\alpha j}/2}- c_{\alpha j}^\dagger e^{-i \phi_{\alpha j}/2}\right ) \:.
\enea 
A $\pi-$Josephson Junction requires  $\Delta $ having opposite signs at U, between  $\ell 1$ and $r N$.   In the gauge in which  $\Delta$ is real, the OP $\Delta $ has to vanish somewhere along the ring and we choose this point to be D with no loss of generality.
As a  first step, to make the approach as simplest as possible,  deep in the  topological phase,  we will adopt  the Kitaev approximation,  $|\Delta|=t$ \cite{Kitaev:2001}all along the chain and  we  choose $\Delta=t$ in the $\ell $  region  and  $\Delta=-t$ in the $r$ region of the ring.  Thus, the chain Hamiltonian becomes: 
\bea
H_{\ell} +  H_{r}=   -i\:  \frac t 2  \sum _{j=1}^{N-1} \left [  \gamma^\ell_{B j} \gamma^\ell_{A j+1}  -    \gamma^r_{A j} \gamma^r_{B j+1}  \right ] \:.
\label{hamo}
\enea
Pictorially, this kind of hybridization can be represented as  in Fig.\ref{fig1}.  Blue (red) circles represent B (A)-type MFs. The  yellow ellipses denote effective strong coupling between nearest neighbor MFs. Were  Eq.\eqref{hamo} the full Hamiltonian, the $\ell$ and $r$ chains would dimerize with opposite phases. 
Four unpaired MFs would appear: two (the red/A ones) located at U and two (the blu/B ones) at D.

To account for the extra interactions $\Gamma$ and $ut$, following Ref. \cite{Alicea:2011},  we refermionize the Hamiltonian  by including $H_U+H_D$ and by rearranging   the MFs at the boundaries. Three effective Dirac Fermions, $d_A$, $d_B$,  $d_{end}$ are required, located at the U  weak link, and three more ones, $f_A$, $f_B$, $f_{end}$, at the D boundary, according to\cite{nota}:
\begin{widetext}
\beq
\begin{array}{l}
H_{eff} =  \frac{i }{2\sqrt{2}}\: t u\: \left [ \:   f_{end} \left ( f_{B} - f _{A}^\dagger  \right ) - h.c.  \right ] +  t \: \left [ d^\dagger _A d_B + h.c. \right ]   \\
- \frac{\Gamma}{2} \left[
\sin \frac \varphi 2 \:  \left (2d^\dagger_{end}d_{end} -1\right ) -i \sqrt  2 \cos \frac \varphi 2 \:  \left (d_{end} (d_B+d^\dagger_A)- h.c.\right ) + \sin\frac \varphi 2 \: \left (d_Bd_A+d^\dagger_A d^\dagger_B+d_A d^\dagger_A-d_Bd^\dagger_B\right )
\right ],
\end{array}\label{Heff} 
\eneq
\end{widetext}
where 
\beq
\varphi = \phi _{\ell1} - \phi_{rN} =  \frac{2e}{\hbar c }�\oint   \vec{A}\cdot d\vec{\ell} \:.
 \eneq
 is the phase difference at the U weak link.  The total energy only depends on the flux threading the ring, $ \Phi = �\oint   \vec{A}\cdot d\vec{\ell} $, in units of  $ \phi _o  $. 
  This Bogolubov-de-Gennes (BdG)  Hamiltonian $H_{eff} $ changes sign under the operation  $\Xi$ of exchanging particles   with holes  and conserves the Fermionic parity 
 $ \mathcal{P} = (-1)^{n} $ (n is the number of fermions at the weak links). 
It can be shown that the low energy spectrum only depends on the first term in the second line, which involves the Dirac fermion
 $d_{end} = \left ( \gamma _{A1}^\ell +i \: \gamma _{AN}^r \right )/2 $, and on terms involving $ f_{end} , f^\dagger _{end}$  
 with $f_{end} = \left ( \gamma _{B1}^r +i \: \gamma _{BN}^\ell \right )/2 $,  which can be obtained by perturbation theory from the first term of the first line.
In addition, residual interactions, not included in $H_{eff}$, can account for finite size effects.
In particular an  extra coupling  $ z_{\alpha}\Gamma  \propto e^{-N} $ ($\alpha=\ell,r$)  arises from  realistic long range interactions between the edge MFs at each chain,  when the simple $t= |\Delta | $\cite{Kitaev:2001} limitation is relaxed. 
 
We are led to  the minimal 4X4 Hamiltonian  in the Majorana representation,  just involving the four relevant MFs:
\beq  
  H_M=i \Gamma  \left ( \begin{array}{cccc} 0 & \sin\frac  \varphi 2  & z_{\ell} & 0 \\ - \sin \frac \varphi  2  & 0 & 0 &  z_{r} \\ -z_{\ell} & 0 & 0 &  u \\ 0 & -z_{r} & -u & 0 \end{array} \right ) ,
\eneq
 in the basis 
$\left [ \gamma _{A1}^\ell , \:  \gamma _{AN}^r ,\: \gamma _{BN}^\ell ,\: \gamma _{B1}^r \right ]$, and we have redifined  $u=ut/\Gamma$.

\section{Hamiltonian spectrum and its low energy approximation}
 \begin{figure}[!htbp]
\begin{center}
\includegraphics[width=0.9\linewidth]{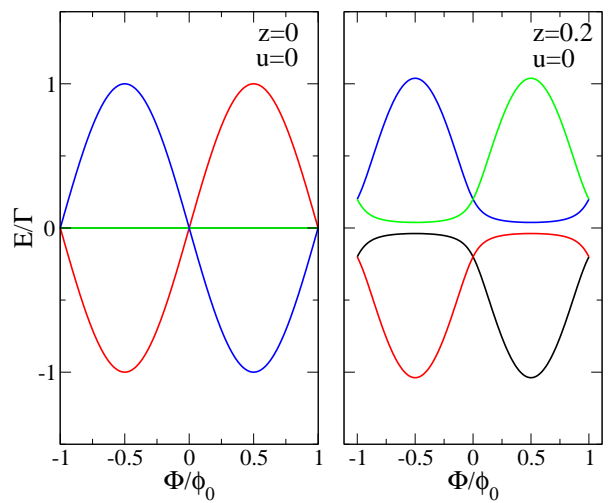}
\caption{(color online)Andreev Bound States as a function of the magnetic flux for an open ring ($u'=0$). Left panel) Finite size corrections not included $z_l=z_r=0$. Right panel) When finite size corrections are included $z_l=z_r=0.2$ there is a splitting of the zero energy Andreev Bound State. }
\label{fig2}
\end{center}
\end{figure}
In this section we show oure results. We first discuss the trivial case in which the ring is open, then we move to the more interesting case of the $\pi$ ring.
\subsection{ Ring cut at the bottom point : $u=0$}
When the control parameter $u$ is set to zero, the ring is cut at the bottom point D and the system is equivalent to a linear topological $\pi$ junction with phase difference $\varphi $ and open ends.
In this case, if the finite size  couplings $z_\alpha$ are neglected (see Fig. \ref{fig2} left panel), there is a crossing at zero energy and zero flux due to the MFs at the U point,  signalling a change of  parity  in the  GS when the flux changes sign.  Together with it,  two dispersionless zero energy modes appear, corresponding to the dangling MFs at the open ends,   However,  the system is expected to be unstable with respect to finite size interactions described by the $z_\alpha$ couplings. Indeed as soon as one turns $z_\alpha$ on, a gap opens and the zero energy MBS disappears (see Fig. \ref{fig2} b).

\subsection{$\pi -$ring configuration: $ u \neq 0 $}
\begin{figure}[htbp]
\begin{center}
\includegraphics[width=0.9\linewidth]{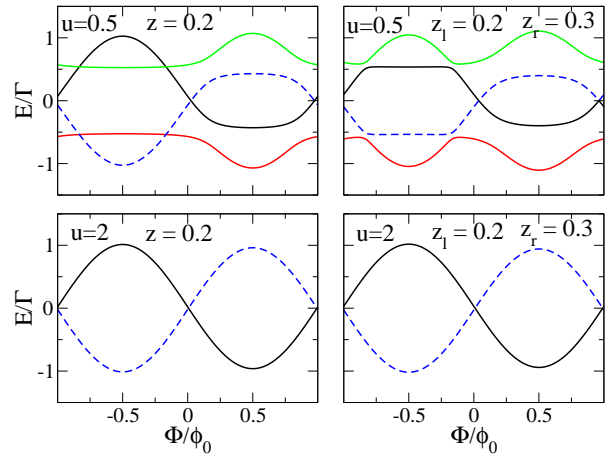}
\caption{(color online) Andreev Bound States as a function of the flux for the $\pi$ ring. Left panels) No asymmetry $z=z_{\ell}=z_{r}$. Right panels) Asymmetrical case: $z_{\ell}\neq z_{r}$. An approximately  symmetric spectrum  is recovered for sizable $u$, independently of the values of the asymmetry parameters $z_{\alpha}$.  In the closed ring geometry the zero energy Majorana Bound State is always present.} 
\label{fig3}
\end{center}
\end{figure}

In the ring configuration ($u\neq0$), the situation is quite different  (see Fig. \ref{fig3}).
The zero energy MBS is always present, no matter  how strong the finite size effects  $z_\alpha$ are.  In the non-symmetric case  ($z_{\ell}\neq z_{r}$), the location of the crossing occurs at non zero flux.  With increasing of $u$, the  flux of the crossing point drifts towards $\varphi =0 $ (see Fig. \ref{fig3} bottom panels).  For the physically relevant case $u>>z_{\ell},z_r$, just the crossing Andreev bound states survive at low energy and the dispersion turns out  to be approximately symmetric. The dispersion of the two crossing low lying energies tends to
 \beq
E_{\pm}  \sim \Gamma \sin\left (\frac \varphi 2\right )  \: \left \langle 2 n_{end} -1 \right \rangle  =  \pm\Gamma \sin \left (\frac \varphi 2\right ).
\label{PiMajo}
\eneq
The minimum is for $ \varphi \approx \pm \pi $ (i.e. flux $ \pm \phi_0 /2 $), depending on the occupancy observable $ n_{end} =  d_{end}^\dagger    d_{end} $ of the MBS located at the Josephson Junction.   

\section{ Model free energy  and stationary conditions of the $rf -$SQUID}
We have shown that,  in the topologically non trivial  $\pi -$ring structure,  an unpaired  zero energy MBS exists and it is robust with respect to pertubations. This  shows up as a crossing of the particle and hole excitation dispersions at flux close to zero. Our  $\pi -$ring modelizes a topologically non trivial  $rf -$SQUID device  and we now argue that the MBS characterizes in a measurable way   the   stationary conditions of the device. 

\begin{figure}[htbp]
\begin{center}
\includegraphics[width=0.9\linewidth]{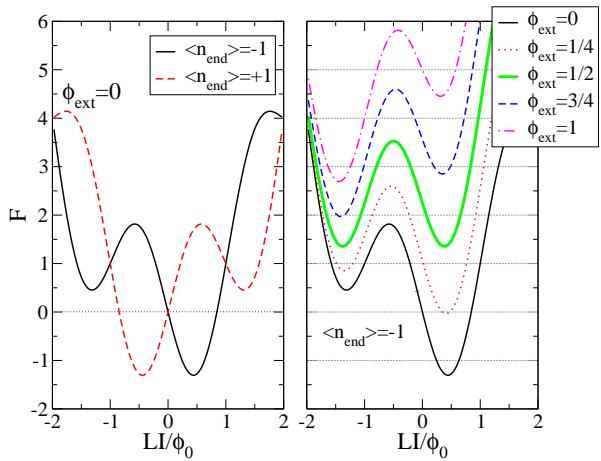}
\caption{(color online) a): free energy vs $LI/\phi_0$ at zero external magnetic field for $\phi_{ext}=0$ ($ u  = 2$ and $ z_\ell = 0.2, z _r =0.3 $). Full (dashed) line  belong to the two  different fermion parity  ${\mathcal{P}}$ . b):  Corresponding  free energy for different values of the external magnetic flux and one single parity. The curves are displaced in energy for clarity}
\label{fig4}
\end{center}
\end{figure}

If the ring has a small diameter, so that its  self-inductance  $L$ cannot be disregarded, the  free energy is a function of the circulating current and of an external flux $\phi _{ext}$ which may be intentionally added. Its simplest form, arising from Eq.\ref{PiMajo} is:    
\beq
F_{\pm} \left (I,\phi_{ext}\right ) = \frac 1 2 L I^2 \pm   \frac{\phi_0 I_c}{ 2 \pi} \sin \left(\frac{ \pi}{\phi_0} (\phi_{ext}+L I )  \right).
\eneq
We have disregarded charging effects that are always negligible in all HTS structures, 
unless the dimensions are scaled to a few hundred nanometers. Charging effects  have been also considered here\cite{vanHeeck:2011}, with no effect on the current periodicity, provided that the entire ring is in a topological nontrivial state.
The  free energy  $ F_{\pm} \left (I,\phi_{ext} = 0\right )$ at zero external flux is plotted in Fig. \ref{fig4}a).   The first and second  minimum,  belonging to the same  \pari $, \: $  differ in phase by $ \approx 4\pi $. Fig.\ref{fig4}b reports the change in shape of the free energy  for one single  \pari, at different applied fluxes $\phi_{ext} $. 

The similarity with the conventional YBCO \fpr is only superficial.  At first sight Fig.  \ref{fig4}b could report the free energy plot of a conventional  \fpr. By fine-tuning the external flux  the two energy minima can be made  degenerate. If   \pari is strictly conserved, this would be the end of the story and no difference would arise.  However,  in the  topologically non trivial case, there is a corresponding set of curves belonging to the other  \pari .  A switching between the two different minima  e.g. at $\phi_{ext} = 0$ (see  Fig.  \ref{fig4}a) not only requires a flux quantum entering or exiting the ring but  a simultaneous  change of   \pari. 

A tool to change the parity could be a side quantum point contact  (QPC) controlling the  charge tunneling. Addition of an electron on the ring would suddenly require the switching  of the whole device between the two possible GSs, corresponding to a jump in the  trapped spontaneous flux.   By contrast, a conventional \fpr  is expected to be widely insensitive to the in-out tunneling of induced charges. Operating with a side gate on the  QPC allows  to distinguish a topological non trivial ring from a conventional one. 

However,  in real life, the job of  fixing \pari appears to be rather hard. The ring is not expected to be isolated: background impurities could provide charge noise, by releasing or capturing charges. For a system open to the environment, the energy spectrum of the isolated ring looses meaning and a description of the state of affairs in terms of the statistical density matrix $\hat \rho (t) $  is required. The latter accounts  for 
the transitions, with absorption and emission of the energy between the two parities and with simultaneous switching of the flux. Under these conditions, the expectation value of the observable corresponding to the flux, is likely to average to zero:
\beq
\lim _{t \to \infty}  \left  \langle \Phi \right \rangle  = {\bf Tr } \left \{  \hat \rho \:  \Phi \right \}  \sim  0 
\eneq

\section{Concluding remarks}

TIs or semiconducting nanowires with strong spin-orbit coupling in proximity  with a singlet superconductor can host MBSs  at their ends as predicted by the Kitaev model\cite{Kitaev:2001}.
There are various experiments involving charge transport, to provide evidence of the presence of the zero energy MBS \cite{Kouwenhoven:2012,Deng:2012,Heiblum:2012,Marcus:2013}.

In this paper we have proposed an experiment which requires a flux measurement.   We believe that the strong point of our proposal is that investigating the thermodynamic equilibrium of an otherwise isolated system rules out some unavoidable doubts and ambiguities which concern the interpretation of transport
measurements\cite{DeFranceschi:2012,Marcus:2013}.
Our  proposal exploits  the unique feature of a d-wave order parameter of the HTS which is used to induce superconductivity, by proximity effect, in a one dimensional conduction channel,  deposited as a ring, on top of a HTS tricrystal structure (see Fig \ref{fig1}).  A ring geometry is highly convenient because magnetic fluxes, generated by circulating currents, can be easily measured  by a scanning-SQUID with high precision.  We have proved  that  any undesired interaction between Majorana quasiparticles, which could lead to the splitting of the zero energy MBS, cannot take
place  in  the proposed ring because of topological protection.
 
Rings etched on conventional tricrystals present weak Josephson coupling at the grain boundaries between differently oriented crystals, and a frustration occur,  provided that $L I_c >>\phi_0$, with a  spontaneous half flux quantum   generated at the center of the ring,  as shown in Ref.\onlinecite{Tsuei:1994}.

\begin{figure}[!htbp]
\begin{center}
\includegraphics[width=0.49\linewidth]{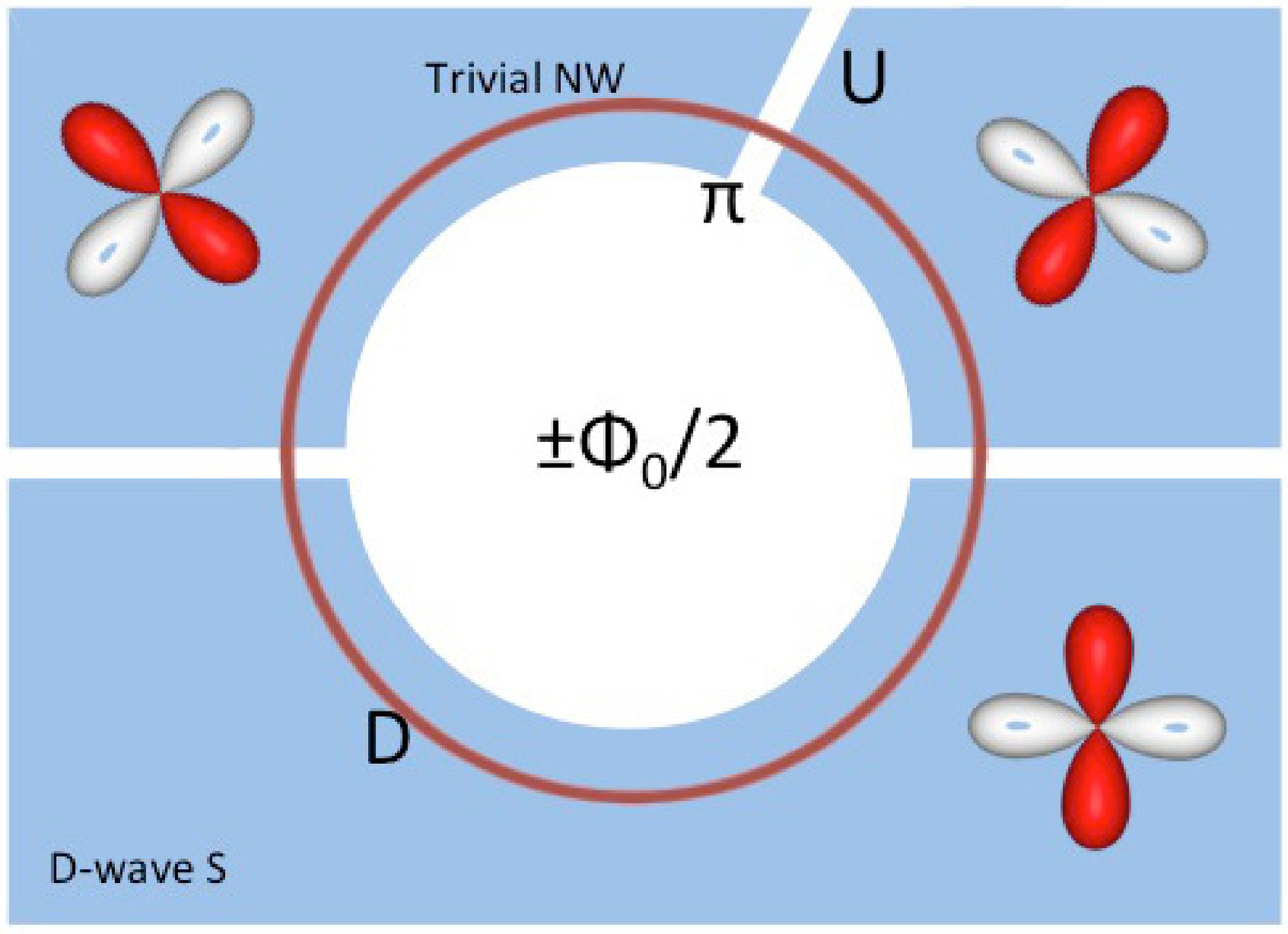}
\includegraphics[width=0.49\linewidth]{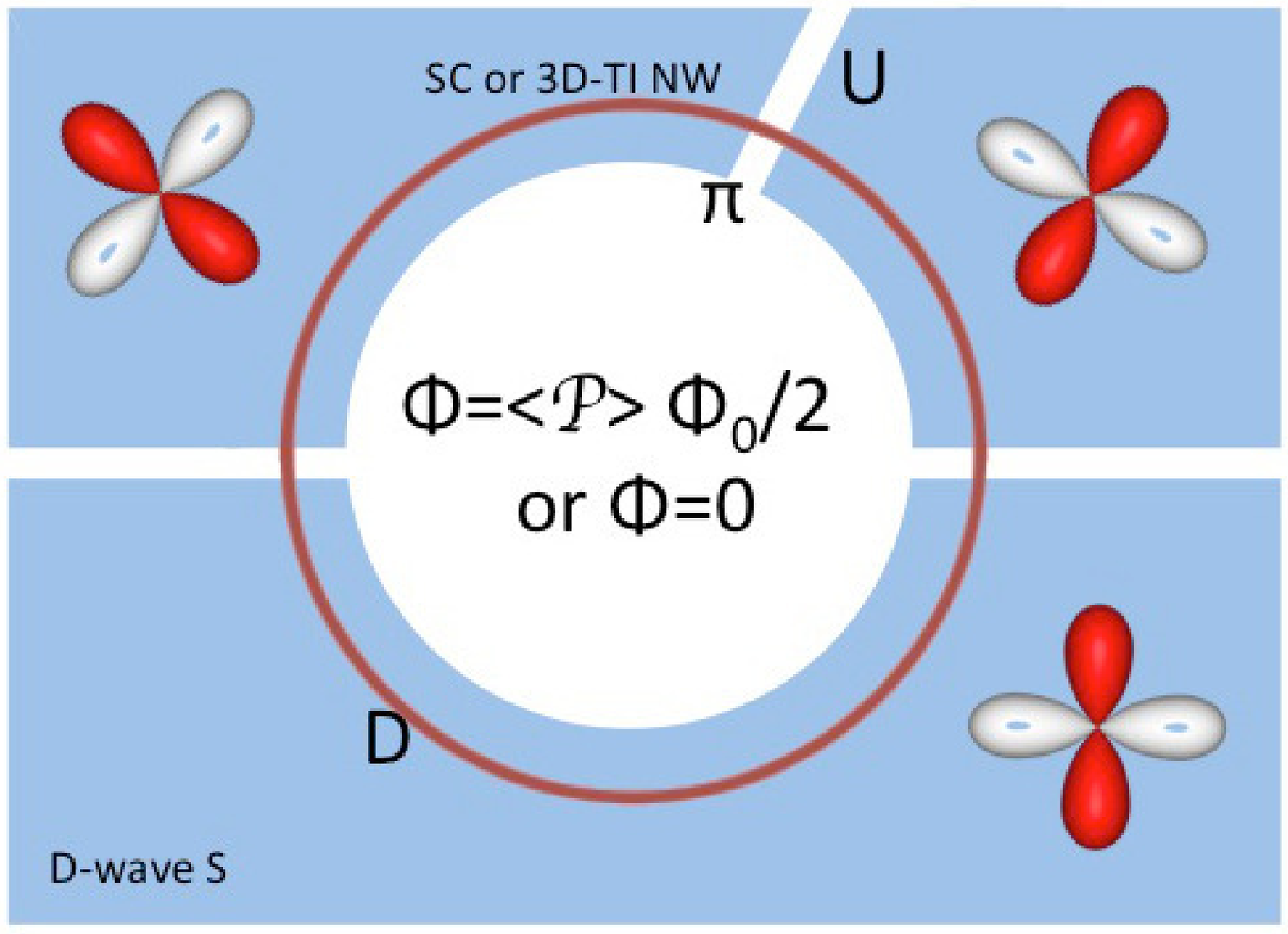}
\caption{(color online) Comparison of the possible spontaneous flux  states of a trivial and a nontrivial pi-ring. }
\label{compara}
\end{center}
\end{figure}

In our case, trenches between the crystals do not allow for Josephson coupling and no flux is trapped at the center of the structure until a ring  is deposited on top of the HTS tricrystal. The three SNS weak links are determined by nanowire barrier and can sustain a supercurrent in the ring. Hence, again frustration can occur, provided that $L I_c >>\phi_0$.
The top  ring can  be either  made of a topological insulator nanowire or by a semiconducting nanowire with strong spin orbit coupling and magnetic field (topological non-trivial $\pi$ ring).
The experiment we have suggested rests  on the possibility of realizing control measurements on the same structure, where the topological insulator or semiconducting nanowire is replaced by a standard metallic nanowire (e.g. Au), thus  which does not display any topological protection (see fig.\ref{compara}).
The spontaneous and stable flux supported by a trivial tricrystal ring is in striking contrast  with what expected in our case.
The free energy of a topologically non trivial ring is depicted in Fig. \ref{fig4}a). It displays two minima, but it is qualitatively different from the the trivial case in that the two branches correspond to different fermion parities and the protected zero energy crossing corresponds to a MBS which can be occupied or empty.
Two different scenarios are possible: if quasiparticle poisoning is negligible, parity is conservered and by changing the number of particles in the ring we can modify the flux state of the $\pi$ ring.
By contrast, in the presence of quasiparticle poisoning, relaxation takes place and the expected flux is zero.
An experiment comparing the behavior of the spontaneous fluxes of two trycristal structures with different rings (one topological and the other
trivial) would give an unambiguous proof of the presence of the  MBS (see Fig. \ref{compara}).
This would make the answer  hopefully sharp and universal.

\begin{acknowledgments}
We acknowledge enlightening discussions with P. Brouwer and A. Golubov. Financial support from FIRB 2012 project "HybridNanoDev" (Grant No.RBFR1236VV),  FP7/2007-2013 (Grant N. 264098) - MAMA  and MIUR  PRIN 2009 project  "Nanowire high critical temperature superconductor field-effect devices"  are gratefully acknowledged.
\end{acknowledgments}

\bibliographystyle{apsrev}

\end{document}